\documentclass[11pt]{article}
\usepackage{amsmath,amsfonts,amssymb,epsfig,wick,a4}

\newcommand{\rightslash}{\! \stackrel{\rightarrow}{\!\slash{\partial}}}
\newcommand{\leftslash}{\! \stackrel{\leftarrow}{\!\slash{\partial}}}
\newcommand{\leftDslash}{\! \stackrel{\leftarrow}{\slash{\! D}}}

\renewcommand\slash[1]{\not \! #1}

\def\Ob{{\mathchoice{\hbox{$\displaystyle\mathbb O$}}
        {\hbox{$\textstyle\mathbb O$}}{\hbox{$\scriptstyle\mathbb O$}}
        {\hbox{$\scriptscriptstyle\mathbb O$}}}}

\newcommand{\be} {\begin{equation}}
\newcommand{\ee} {\end{equation}}
\newcommand{\bma} {\begin{math}}
\newcommand{\ema} {\end{math}}
\newcommand{\beqa} {\begin{eqnarray}}
\newcommand{\eeqa} {\end{eqnarray}}

\newcommand{\ga} {\gamma}

\newcommand{\sig} {\sigma}
\newcommand{\nn} {\nonumber}
\newcommand{\bc} {\begin{center}}
\newcommand{\ec} {\end{center}}

\newcommand{\simgt}{\hbox{ \raise3pt\hbox to 0pt{$>$}
    \raise-3pt\hbox{$\sim$} }}
\newcommand{\simsm}{\hbox{ \raise3pt\hbox to 0pt{$<$}
    \raise-3pt\hbox{$\sim$} }}
\begin{document}

\begin{titlepage}
\begin{flushright}
ECT$^*$-06-10\\
HD-THEP-06-16\\
hep-ph/0608082
\\
\end{flushright}
\vfill
\begin{center}
\boldmath
{\LARGE{\bf Chiral Symmetry and Diffractive Neutral Pion}}\\[.2cm]
{\LARGE{\bf Photo- and Electroproduction}}
\unboldmath
\end{center}
\vspace{1.2cm}
\begin{center}
{\bf \Large
Carlo Ewerz\,$^{a,1}$, Otto Nachtmann\,$^{b,2}$
}
\end{center}
\vspace{.2cm}
\begin{center}
$^a$
{\sl
ECT\,$^*$, Strada delle Tabarelle 286, 
I-38050 Villazzano (Trento), Italy}
\\[.5cm]
$^b$
{\sl
Institut f\"ur Theoretische Physik, Universit\"at Heidelberg\\
Philosophenweg 16, D-69120 Heidelberg, Germany}
\end{center}                                                                
\vfill
\begin{abstract}
\noindent
We show that diffractive production of a single neutral pion 
in photon-induced reactions at high energy is dynamically 
suppressed due to the approximate chiral symmetry of QCD. 
These reactions have been proposed as a test of the odderon 
exchange mechanism. 
We show that the odderon contribution to the amplitude for 
such reactions vanishes exactly in the chiral limit. 
This result is obtained in a nonperturbative framework and by 
using PCAC relations between the amplitudes for neutral 
pion and axial vector current production. 
\vfill
\end{abstract}
\vspace{5em}
\hrule width 5.cm
\vspace*{.5em}
{\small \noindent
$^1$ email: Ewerz@ect.it \\
$^2$ email: O.Nachtmann@thphys.uni-heidelberg.de
}
\end{titlepage}

\section{Introduction}
\label{sec:intro}

\begin{figure}[ht]
\begin{center}
\vspace*{.4cm}
\includegraphics[width=7.5cm]{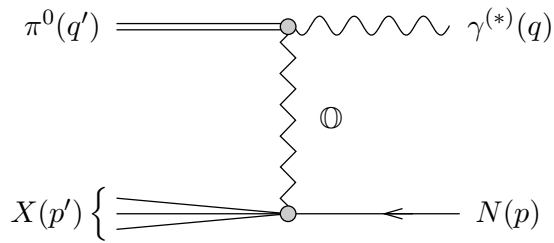}
\end{center}
\caption{Diffractive production of a neutral pion in real or virtual 
photon-nucleon scattering due to exchange of an odderon ($\Ob$). 
\label{figure1}}
\end{figure}

In this paper we study the diffractive production of a single neutral pion 
in the scattering of a real or virtual photon on a nucleon:
\be\label{1}
\gamma^{(*)}(q)+N(p) \, \longrightarrow \, \pi^0(q')+X(p')\,.
\ee
Here $N$ stands for a proton or a neutron, and $X$ denotes the  rest of the 
hadronic final state which can consist of a single nucleon or of a group of 
hadrons. The four-momenta are indicated in brackets. The usual 
invariant variables are 
\be\label{1a}
s=(p+q)^2=(p'+q')^2\,,\quad \quad \quad 
t=(p-p')=(q-q')^2 \,.
\ee
We always consider high energies, that is $s\gg m^2_p$. 
We assume that there is a large rapidity gap between $\pi^0$ and $X$ 
in (\ref{1}). Since neutral pions have charge conjugation $C=+1$ these 
reactions are at high energy expected to occur due to the exchange 
of an odderon, the $C=-1$ partner of the well-established pomeron, 
see figure \ref{figure1}. (Note that throughout this paper we 
draw the incoming particles to the right.) For a review of 
high energy scattering in QCD see \cite{Donnachie:en}. 

The odderon was introduced many years ago 
\cite{Lukaszuk:1973nt,Joynson:1975az} 
and since then has been studied extensively from the theoretical 
point of view, for a review see \cite{Ewerz:2003xi}. Recently, 
particular progress has been made in understanding the odderon 
in the perturbative regime \cite{Janik:1998xj,Bartels:1999yt}. 
From the experimental side the odderon has turned out to be an elusive 
object. There is some evidence for it in high energy proton-proton and 
antiproton-proton scattering \cite{Breakstone:1985pe} 
at a momentum transfer squared of $|t|\approx 1.0-1.5 \,\mbox{GeV}^2$, 
see also \cite{Dosch:2002ai} for a recent discussion. 
But otherwise conclusive evidence for the existence of the odderon is missing. 
In \cite{Schafer:1992pq,Barakhovsky:1991ra,Kilian:1997ew} 
it was suggested to look for the odderon in the reaction (\ref{1}). 
Subsequently, the photoproduction of $\pi^0$ was investigated in detail in 
\cite{Berger:1999ca}. (For a discussion of the photoproduction of tensor 
and other pseudoscalar mesons see \cite{Berger:2000wt,Kilian:1997ew}.) 
The cross section at a c.\,m.\ energy of $\sqrt{s}=20$ GeV was predicted 
to be  
\be \label{2}
\sig(\ga p \to \pi^0 X) \approx 300 \,{\rm nb} \,,
\ee
and only a weak dependence of this result on the energy $\sqrt{s}$ 
is expected. The uncertainties of the model of nonperturbative dynamics 
used in \cite{Berger:1999ca} imply a rather large uncertainty of about 
a factor $2$ in the prediction (\ref{2}). 
However, corresponding experimental searches at HERA found no evidence 
for odderon-exchange reactions. 
The experimental search at $\sqrt{s} = 200\,\mbox{GeV}$ \cite{Adloff:2002dw} 
resulted in an upper limit of 
\be\label{3}
\sigma(\ga p \to \pi^0 N^*)  <  49\,{\rm nb}
\ee 
at the 95\,\% confidence level, hence excluding the prediction (\ref{2}) 
even if the large uncertainty inherent in the latter is taken into account. 
The non-observation of diffractive single pion production at HERA is 
especially striking since among all reactions in which hadrons are 
diffractively produced this reaction is the one with the largest kinematical 
phase space. Therefore there must be a dynamical mechanism which 
strongly suppresses the production rate. 

In a short paper \cite{Donnachie:2005bu} possible causes for the failure 
of the calculations of \cite{Berger:1999ca} in comparison with experiment 
were discussed. One of them is a very low odderon intercept leading to a 
strong suppression of the cross section for the process (\ref{1}) at high 
energies. A second possibility discussed in \cite{Donnachie:2005bu} is the 
failure of a factorisation hypothesis for field strength correlators that 
had been used in the nonperturbative model underlying the calculation 
of \cite{Berger:1999ca}. Another known source of suppression of 
odderon-induced reactions is the potentially small coupling of the 
odderon to the nucleon. A possible reason for the smallness of this 
coupling is a clustering of two constituent quarks of the nucleon 
into a small-size system of diquark type \cite{Rueter:1996yb,Dosch:2002ai}. 
However, the suppression 
due to that effect should only be relevant for reactions in which 
the proton stays intact, but should not lead to a sizable effect 
in reactions of type (\ref{1}) 
in which the proton dissociates or is excited \cite{Rueter:1998gj}. 

Finally, it was pointed out in \cite{Donnachie:2005bu} that a suppression 
of the cross section for diffractive single pion production 
can occur due to the particular properties of the wave function of 
the pion. These were not properly taken into account in the 
calculation leading to the prediction of \cite{Berger:1999ca}. 
It is in fact natural to expect that 
the special nature of the pion in the context of chiral symmetry can 
have considerable effects on the reaction (\ref{1}). 

In the present paper we give a detailed account of the latter argument. 
We shall show that the chiral symmetry of QCD leads to a vanishing 
amplitude for the reaction (\ref{1}) when the limits of high 
energies and vanishing pion mass are taken. 
Our analysis is entirely based on nonperturbative techniques. In 
particular, we shall use the functional methods explained in detail in 
\cite{Ewerz:2004vf,Ewerz:2006vd} in connection with the dipole 
picture for photon-induced reactions. 

Our paper is organised as follows. In section \ref{sec:react} 
we discuss $\pi^0$-production in a functional integral approach and 
use PCAC to relate this reaction to one involving the axial vector current. 
In section \ref{sec:classification} we classify the contributions to the 
amplitude at high energies and identify the contributions which are 
leading at high energies. In section \ref{sec:divrel} we find 
the dependence of the latter on the light quark masses. In section 
\ref{sec:renormalisation} we study the renormalisation of the 
amplitudes under consideration and find that the dependence on 
the light quark masses remains unchanged. In section \ref{sec:results} 
we finally show that the leading terms at high energies in fact vanish in 
the chiral limit $m_\pi\rightarrow 0$, and we discuss this result. 
In appendix \ref{sec:appA} we 
describe the functional methods used in section \ref{sec:classification}. 
In appendix \ref{sec:appB} we outline how our results can be 
generalised to single diffractive pion production with a break-up 
of the nucleon. 

\boldmath
\section{The reactions $\gamma^{(*)}p\rightarrow \pi^0 p$ and 
$\gamma^{(*)}p\rightarrow A^3 p$}
\unboldmath
\label{sec:react}

In this section we consider as an example for (\ref{1}) neutral pion 
photo- and electroproduction on a proton
\be\label{5}
\gamma^{(*)}(q,\nu)+p(p,s)\, \longrightarrow \,\pi^0(q')+p(p',s') \,.
\ee
Momenta and spin labels are indicated in brackets. We suppose
\be\label{6}
q^2=-Q^2\le 0
\ee
and have for real pions
\be\label{7}
q'^2=m^2_\pi \,.
\ee

Let $\phi^a(x)$ be a renormalised and correctly normalised 
interpolating field operator for the isotriplet of pions, $a=1,2,3$. 
We have then
\be\label{8}
\langle 0|\phi^a(x)|\pi^b(q')\rangle=e^{-iq'x}\delta_{ab}\,,
\ee
and the physical pion states are 
\beqa\label{9}
|\pi^0(q')\rangle &=&{} |\pi^3(q')\rangle,\nn \\
|\pi^\pm(q')\rangle &=&{} \frac{1}{\sqrt{2}}
\left(|\pi^1(q')\rangle\pm i\,|\pi^2(q')\rangle\right)\,.
\eeqa
The LSZ reduction formula \cite{Lehmann:1954rq} gives for the 
amplitude of (\ref{5}) 
\beqa\label{10}
\lefteqn{(2\pi)^4\delta^{(4)}(p'+q'-p-q) \,{\cal M}^\nu_{s's}(\pi^0;q',p,q)}
\nn\\
&=&{}-i \int d^4x' \,d^4x \, e^{iq'x'}e^{-iqx}
\left(\Box_{x'}+m^2_\pi\right) \,\langle p(p',s')|{\rm T}^*\phi^3(x')
J^\nu(x)|p(p,s)\rangle\,.
\eeqa
Here $e J^\nu(x)$ is the hadronic part of the electromagnetic current 
with $e$ the proton charge. With the quark field operator 
\be\label{10a}
\psi(x)=\left(\begin{array}{l}
u(x)\\d(x)\\s(x)\\c(x)\\b(x)\\t(x)\end{array}\right)
\ee
we have
\be\label{10b}
J^\nu(x)=\bar{\psi}(x)\gamma^\nu \mathbf{Q}\psi(x)\,.
\ee
Here $\mathbf{Q}=$ diag$(Q_u,\dots,Q_t)$ is the quark charge matrix 
with $Q_u=2/3,~Q_d=-1/3$ etc. Our normalisation is such that the 
$T$-matrix element for (\ref{5}) with a real photon of polarisation 
vector $\varepsilon^\nu$ is given by
\be\label{11}
\langle \pi^0(q'),p(p',s')|{\cal T}|\gamma(q,\varepsilon),p(p,s)\rangle
=e {\cal M}^\nu_{s's}(\pi^0;q',p,q) \,\varepsilon_\nu\,.
\ee
In (\ref{11}) we have, of course, $q^2=0$ and $q^{\prime 2}=m^2_\pi$. 
Our conventions for kinematics, Dirac matrices etc.\ follow \cite{19a}.

Starting from (\ref{10}) we can extend the amplitude ${\cal M}^\nu_{s's}$ 
to off-shell pions, that is, we consider in the following 
${\cal M}^\nu_{s's}(\pi^0;q',p,q)$ of (\ref{10}) for
\be\label{12}
q^{\prime 2}\le m^2_\pi\,,\quad\quad\quad q^2=-Q^2\le 0\,.
\ee

The reaction which we consider along with (\ref{5}) is 
\be\label{16}
\gamma^{(*)}(q,\nu)+p(p,s) \, \longrightarrow \, A^3(q',\mu)+p(p',s')\,,
\ee
that is, the production of an axial vector current instead of the $\pi^0$ 
meson. The isotriplet of axial vector currents is given by 
\be\label{13a}
A^a_\mu(x)=\bar{\psi}(x)\gamma_\mu\gamma_5 \mathbf{T}_a \psi(x) 
\quad \quad
(a=1,2,3)\,.
\ee
Here we denote by 
\be\label{14a}
\mathbf{T}_a=\left(\begin{array}{c|c}
\frac{1}{2}\tau_a&0 \\ \hline0&0
\end{array}\right)
\ee
the flavour isospin matrices for the quarks, where the $\tau_a$ are the 
Pauli matrices.
We define the amplitude for reaction (\ref{16}) as
\beqa\label{17}
\lefteqn{
(2\pi)^4\delta^{(4)}(p'+q'-p-q)\,
{\cal M}^{\mu\nu}_{s's}(A^3;q',p,q)
}
\nn\\
&=&{}\frac{i}{2\pi m_p}\int d^4x'\,d^4x \,e^{iq'x'} e^{-iqx}
\langle p(p',s')|\,{\rm T}^* A^{3\mu}(x')J^\nu(x)|p(p,s)\rangle\,.
\eeqa
For (\ref{17}) we consider again the kinematic region (\ref{12}). 

The well known PCAC relation (partially conserved axial vector current) 
relates the divergence of the currents (\ref{13a}) to a correctly normalised 
pion field operator, 
\be\label{15}
\partial_\lambda A^{a\lambda}(x)=
\frac{f_\pi m^2_\pi}{\sqrt{2}} \, \phi^a(x)\,,
\ee
see for example \cite{AdlerDashen}. 
Here $f_\pi\cong130\,\mbox{MeV}$ is the pion decay constant, 
see p.\ 496 of \cite{Eidelman:2004wy}. 

We insert now the PCAC relation (\ref{15}) in (\ref{10}) and get 
for $q^{\prime 2}<m^2_\pi$ 
\beqa\label{18}
\lefteqn{
(2\pi)^4\delta^{(4)}(p'+q'-p-q) \,{\cal M}^\nu_{s's}(\pi^0;q',p,q)
}
\\
&=&{}-i \int d^4x'\,d^4x~e^{iq'x'} e^{-iqx}
\frac{\sqrt{2}}{f_\pi m^2_\pi}\,(-q^{\prime 2}+m^2_\pi)
\langle p(p',s')|\,{\rm T}^*\partial'_\mu A^{3\mu}(x')J^\nu(x)|p(p,s)\rangle\,.
\nn
\eeqa
An integration by parts and using the vanishing of the equal-time commutator 
\be\label{19}
\left[A^{30}(x'),J^\nu(x)\right]\,\delta(x^{\prime 0}-x^0)=0
\ee
leads to 
\be\label{20}
{\cal M}^\nu_{s's}(\pi^0;q',p,q)=
\frac{2\pi m_p\sqrt{2}}{f_\pi m^2_\pi}\,
(-q'^2+m^2_\pi) \,iq'_\mu {\cal M}^{\mu\nu}_{s's}
(A^3;q',p,q) \,,
\ee
 or, written differently, 
\be\label{21}
iq'_\mu{\cal M}^{\mu\nu}_{s's}(A^3;q',p,q)=
-\frac{f_\pi m^2_\pi}{2\pi m_p\sqrt{2}}\,
\frac{1}{q'^2-m^2_\pi +i\epsilon}\,
{\cal M}^\nu_{s's}(\pi^0;q',p,q)\,.
\ee

\begin{figure}[ht]
\begin{center}
\vspace*{.4cm}
\includegraphics[width=10cm]{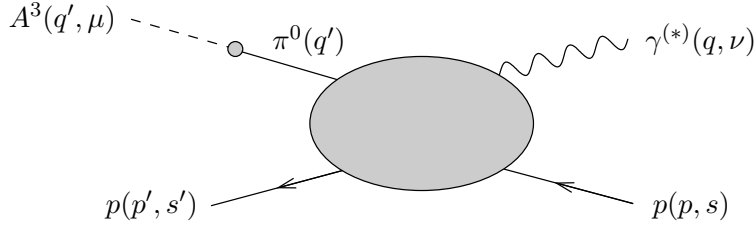}
\end{center}
\caption{Pion-pole contribution to the amplitude 
${\cal M}^{\mu\nu}_{s's}(A^3;q',p,q)$ in (\ref{17}). \label{figure2}}
\end{figure}

Let us as a side remark remind the reader at this point that taking 
the limit $q'^2\rightarrow 0$ in (\ref{21}) leads to a Goldberger-Treiman 
type relation \cite{Goldberger:1958vp}. 
Indeed, we can split the amplitude 
${\cal M}^{\mu\nu}_{s's}(A^3;q',p,q)$ into the pion pole part (see 
figure \ref{figure2}) and the rest which has no pion pole. 
The pole term must be proportional to $q'^\mu$ and its residue 
is fixed by (\ref{20}). We can take the pole term to be such that 
\beqa\label{22}
{\cal M}^{\mu\nu}_{s's}(A^3;q',p,q)
&=&{}
i\,\frac{f_\pi}{2\pi m_p\sqrt{2}}\,
\frac{q'^\mu}{q'^2-m^2_\pi+i\epsilon} \,{\cal M}^\nu_{s's}(\pi^0;q',p,q)
\nn\\
&&{}
+\left. {\cal M}^{\mu\nu}_{s's}(A^3;q',p,q)\right|_{\textup{non-pole}}\,.
\eeqa
Inserting this in (\ref{21}) leads to
\beqa\label{23}
\frac{f_\pi}{2\pi m_p\sqrt{2}}\,
\frac{(-m^2_\pi)}{q'^2-m^2_\pi+i\epsilon}\,
{\cal M}^\nu_{s's}(\pi^0;q',p,q)
&=&{}
\frac{f_\pi}{2\pi m_p\sqrt{2}}\,
\frac{(-q'^2)}{q'^2-m^2_\pi+i\epsilon}\,
{\cal M}^\nu_{s's}(\pi^0;q',p,q)
\nn \\
&&{}
+\left. 
iq'_\mu{\cal M}^{\mu\nu}_{s's}(A^3;q',p,q)\right|_{\textup{non-pole}}
\,.
\eeqa
Taking now the limit $q'^2\rightarrow 0$ in (\ref{21}) and (\ref{23}) we 
get the Goldberger-Treiman type relation
\beqa\label{24}
\left. {\cal M}^\nu_{s's}(\pi^0;q',p,q)\right|_{q'^2=0}
&=&{}
\frac{2\pi m_p\sqrt{2}}{f_\pi}\,
iq'_\mu \left.{\cal M}^{\mu\nu}_{s's}(A^3;q',p,q)\right|_{q'^2=0}
\nn \\
&=&{}
\frac{2\pi m_p\sqrt{2}}{f_\pi}\,
iq'_\mu \left. {\cal M}^{\mu\nu}_{s's}
(A^3;q',p,q)\right|_{\textup{non-pole},\,q'^2=0}.
\eeqa
Note that at $q'^2=0$ the pion pole amplitude on the r.h.s.\ of (\ref{23}) 
gives no contribution. The Goldberger-Treiman type relations are between 
the pion amplitude extrapolated to $q'^2=0$ and the current amplitude 
$q'_\mu {\cal M}^{\mu\nu}_{s's}(A^3;q',p,q)$ at $q'^2=0$, where only 
the non-pole term contributes.

\boldmath
\section{Classification of diagrams for $\gamma^{(*)} p\rightarrow A^3  p$}
\label{sec:classification}
\unboldmath

\begin{figure}[ht]
\begin{center}
\vspace*{.4cm}
\includegraphics[width=14.45cm]{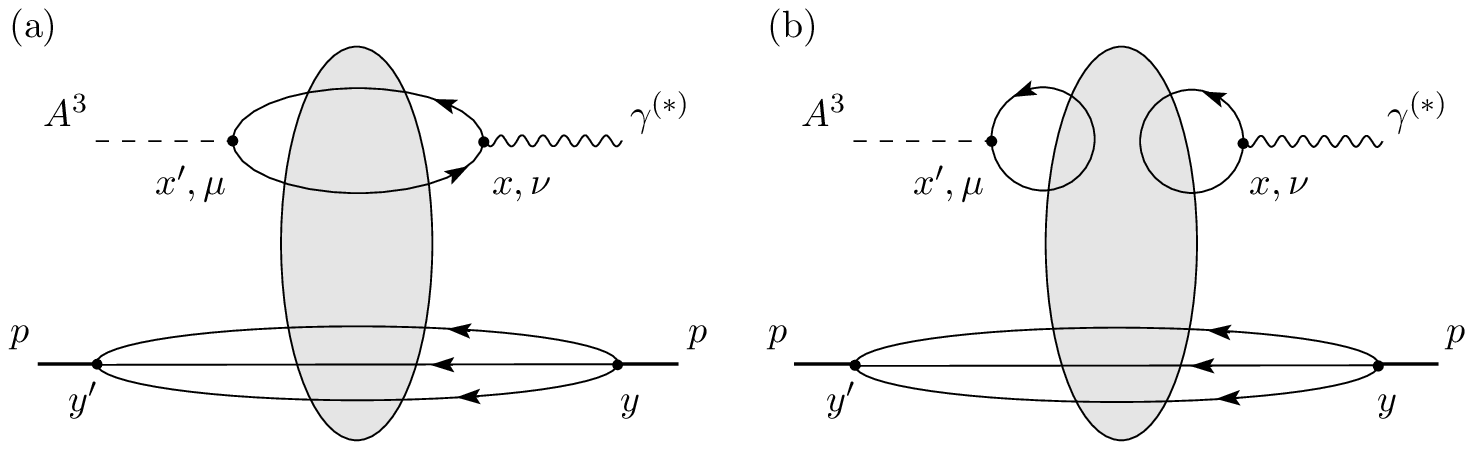}
\end{center}
\caption{Diagrams which are expected to be the leading ones for 
reaction (\ref{16}) at high energies. \label{figure3}}
\end{figure}

In \cite{Ewerz:2004vf} we discussed real and virtual Compton scattering, 
$\gamma^{(*)}p\rightarrow\gamma^{(*)}p$, using functional methods. 
In particular, we gave a classification of contributions to the amplitude in 
terms of nonperturbative diagrams and identified the diagram classes 
which should be the leading ones at high energies; see section 2 
of \cite{Ewerz:2004vf}. The general classification scheme into diagram 
classes (a) to (g) of figure 2 in \cite{Ewerz:2004vf} holds 
unchanged for reaction (\ref{16}). All we have to do is to replace the 
electromagnetic current representing the final state photon in the 
Compton scattering case by the axial vector current. Most of the discussion 
of which nonperturbative diagrams are expected to be leading at high 
energies can be taken over from section 2.2 of \cite{Ewerz:2004vf}. 
There are diagrams with pure multi-gluon exchange in the $t$-channel 
as shown in figure \ref{figure3}a and \ref{figure3}b. 
In our case this exchange must have odderon 
quantum numbers, that is $C=-1$. The analogues of the diagrams (c) to (g) of 
figure 2 of \cite{Ewerz:2004vf} for our case correspond to quark exchange in 
the $t$-channel. As explained in section 2.2 of \cite{Ewerz:2004vf} such 
diagrams are expected to be suppressed for large $s$. 
Thus, the diagrams interesting for the odderon search are those shown 
in figure \ref{figure3}a and \ref{figure3}b. 
As explained in detail in 
\cite{Ewerz:2004vf} the full lines in figure \ref{figure3} 
represent quark propagators in a fixed gluon potential. The shaded blobs 
indicate the functional integral over all gluon potentials with the measure 
given in (\ref{A.8}) in appendix \ref{sec:appA}. 
As in (13) of \cite{Ewerz:2004vf} we can now write 
\be\label{25}
{\cal M}^{\mu\nu}_{s's}(A^3;q',p,q)=
{\cal M}^{(a)\mu\nu}_{s's}(A^3;q',p,q)+ \ldots 
+{\cal M}^{(g)\mu\nu}_{s's}(A^3;q',p,q)\,,
\ee
according to the decomposition of the amplitude in the diagram 
classes (a) to (g). The relevant terms for us here are
\beqa\label{26}
{\cal M}^{(a)\mu\nu}_{s's}(A^3;q',p,q)
&=&{}
\left\langle U_{s's}(p',p)\,A^{\mu\nu}(q',q)\right\rangle_G\,,
\\
\label{27}
{\cal M}^{(b)\mu\nu}_{s's}(A^3;q',p,q)
&=&{}
\left\langle U_{s's}(p',p) \right. 
\tilde{B}^\mu(q')
\left. B^\nu(q)\right\rangle_G\,.
\eeqa
More details are given in appendix \ref{sec:appA}. 
The physical interpretation and analytic 
expressions of the terms in (\ref{26}) and (\ref{27}) are as follows. 
The scattering amplitude for the $\gamma^{(*)}$ converting to 
$A^{3\mu}$ in the fixed gluon potential $G$ (see the upper part of 
figure \ref{figure3}a) is given by
\be\label{28}
A^{\mu\nu}(q',q)=\int d^4x'\, d^4x\,e^{iq'x'}e^{-iqx}
{\rm Tr}\left[
\gamma^\mu\gamma_5\mathbf{T}_3 \mathbf{S}_F(x',x;G)
\gamma^\nu \mathbf{Q} \,\mathbf{S}_F(x,x';G)\right]\,.
\ee
Here 
\be\label{29}
\mathbf{S}_F(x,x';G)=
\mathrm{diag}\left(S^{(u)}_F(x,x';G),S^{(d)}_F(x,x';G),\dots,
S^{(t)}_F(x,x';G)\right)
\ee
is the propagator matrix for the quarks moving in the fixed gluon 
potential $G$. The factor $B^\nu(q)$ in (\ref{27}) represents the absorption 
of the photon in the fixed gluon potential (see figure \ref{figure3}b), 
\be\label{30}
B^\nu(q)=
\int d^4x \,e^{-iqx}i\,\mathrm{Tr} \left[\gamma^\nu
\mathbf{Q}\,\mathbf{S}_F(x,x;G)\right]\,.
\ee
Similarly, the factor $\tilde{B}^\mu (q')$ in (\ref{27}) represents the 
creation of the axial vector current in the fixed gluon potential 
(see figure \ref{figure3}b), 
\be\label{31}
\tilde{B}^\mu(q')=\int d^4x' \,e^{iq'x'}
i \,\mathrm{Tr}\left[\gamma^\mu\gamma_5\mathbf{T}_3
\mathbf{S}_F(x',x';G)\right]\,.
\ee
The factor $U_{s's}(p',p)$ in (\ref{26}) and (\ref{27}) represents the 
scattering of the proton in the fixed gluon potential. It is given explicitly 
in (\ref{A.11}) in appendix \ref{sec:appA}, 
together with the expression for the functional integral 
$\langle\cdot\rangle_G$. 

\section{Divergence relations for axial vector amplitudes}
\label{sec:divrel}

In this section we study the divergences of the axial vector amplitudes 
in (\ref{26}) and (\ref{27}), that is $q'_\mu A^{\mu\nu}(q',q)$ and 
$q'_\mu\tilde{B}^\mu(q')$. For $A^{\mu\nu}(q',q)$ we find from 
(\ref{28}) with (\ref{A.6}) and (\ref{A.7})
\beqa\label{32}
q'_\mu A^{\mu\nu}(q',q)&=&{} 
i \int d^4x' \, d^4x\,e^{iq'x'} e^{-iqx}
\frac{\partial}{\partial x'^\mu}\,
\mathrm{Tr}\left[\gamma^\mu\gamma_5\mathbf{T}_3 
\mathbf{S}_F(x',x;G)
\gamma^\nu\mathbf{Q}\,\mathbf{S}_F(x,x';G)\right]
\nonumber\\
&=&{}
-\frac{1}{3}\int d^4x'\,d^4x\,e^{iq'x'} e^{-iqx}
\nonumber\\
&&{}
\left\{
2 \,m^{(0)}_u \,\mathrm{Tr}\left[
S^{(u)}_F(x,x';G) \gamma_5 S^{(u)}_F(x',x;G)\gamma^\nu
\right]
\right.
\nonumber\\
&&{}
\left. {}
+m^{(0)}_d \,\mathrm{Tr}\left[
S^{(d)}_F(x,x';G) \gamma_5 S^{(d)}_F(x',x;G)\gamma^\nu
\right]
\right\}
\,.
\eeqa
We see explicitly here that $q'_\mu A^{\mu\nu}(q',q)$ contains one factor 
of the small $u$- and $d$-quark masses and inserting this in (\ref{26}) we 
find the same for $q'_\mu {\cal M}^{(a)\mu\nu}_{s's}(A^3;q',p,q)$. 
With (\ref{20}) this implies that also ${\cal M}^{(a)\nu}_{s's}(\pi^0;q',p,q)$ 
contains one factor of $m_u$ or $m_d$. 
But as we shall see below this factor of the light quark masses 
is cancelled by the factor $m^2_\pi$ in the denominator in (\ref{20}). 
The crucial observation which we will make in the present section 
is that $q'_\mu A^{\mu\nu}(q',q)$ is, in fact, proportional to the square 
of the light quark masses. 

In order to trace the factors of the light quark masses in our amplitudes 
we will in the following make the dependence on $m_q^{(0)}$ explicit. 
We therefore indicate the dependence of the quark propagator 
on the quark mass by an additional argument, 
\be
\label{4tharg}
S^{(q)}_F(x,x';G) = S^{(q)}_F(x,x';G, m^{(0)}_q) \,.
\ee
We further introduce the free propagator for massless quarks, 
\be\label{36}
S^{(q)}_F(x,y;0,0)=-\int\frac{d^4p}{(2\pi)^4}\,e^{-ip(x-y)}\,
\frac{1}{\slash{p} +i\epsilon}\,, 
\ee
which satisfies (\ref{A.6}) and (\ref{A.7}) for $G=0$ and $m^{(0)}_q=0$. 
From the defining equations (\ref{A.6}) and (\ref{A.7}) 
for the full propagator we can easily derive the Lippmann-Schwinger relation 
\be
S^{(q)}_F(x,y;G,m^{(0)}_q)= S^{(q)}_F(x,y;G,0) 
- \int d^4z\, S^{(q)}_F(x,z;G,0) \,m^{(0)}_q S^{(q)}_F(z,y;G,m^{(0)}_q)
\,.
\ee
In matrix notation where the 
space-time arguments and integrations are suppressed this reads 
\be\label{34}
S^{(q)}_F(G,m^{(0)}_q)=
S^{(q)}_F(G,0)-S^{(q)}_F(G,0)\,m^{(0)}_qS^{(q)}_F(G,m^{(0)}_q)\,.
\ee
Similarly, we find for the massless propagator the 
Lippmann-Schwinger relation 
\be\label{35}
S^{(q)}_F(G,0)=
S^{(q)}_F (0,0) 
-S^{(q)}_F (0,0) \,g^{(0)} \! \slash{G}^a \frac{\lambda_a}{2}S^{(q)}_F(G,0)
\,.
\ee
Here $g^{(0)}$ is the unrenormalised QCD coupling constant, and 
$\lambda_a$ are the Gell-Mann matrices.  
From (\ref{35}) we get (still using matrix notation) 
\beqa \label{37}
S^{(q)}_F(G,0) &=&{}
\left[1+S^{(q)}_F(0,0) \,g^{(0)}\! \slash{G}^a\frac{\lambda_a}{2}\right]^{-1}
S^{(q)}_F (0,0)
\nn \\
&=&{}
\sum^\infty_{n=0}
\left(-S^{(q)}_F (0,0) \,g^{(0)}\! \slash{G}^a\frac{\lambda_a}{2}\right)^n 
S^{(q)}_F (0,0) \,.
\eeqa
Since all terms on the r.h.s.\ of (\ref{37}) have an odd number of $\gamma$ 
matrices we find immediately
\be\label{38}
S^{(q)}_F(x,x';G,0)\gamma_5+\gamma_5 S^{(q)}_F(x,x';G,0)=0 \,.
\ee
That is, the massless quark propagator in a fixed gluon potential 
anticommutes with $\gamma_5$. 

\begin{figure}[ht]
\begin{center}
\vspace*{.4cm}
\includegraphics[width=12cm]{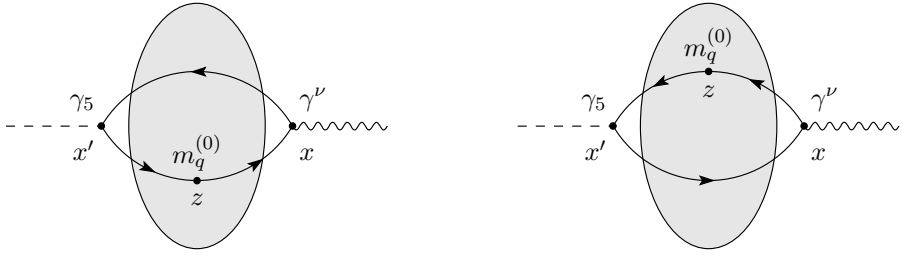}
\end{center}
\caption{Diagrammatic representation of 
$m_q^{(0)} E'^\nu (x,x';G,m^{(0)}_q)$ (\ref{40}) as the sum of two 
quark loops in a given gluon potential. \label{figure4}}
\end{figure}

Let us now consider for $q=u,d$ the trace part of the integrand in (\ref{32}), 
\be\label{33}
E^\nu(x,x';G,m^{(0)}_q)
=\mathrm{Tr}\left[
S^{(q)}_F(x,x';G,m^{(0)}_q)\gamma_5  S^{(q)}_F(x',x;G,m^{(0)}_q)\gamma^\nu 
\right]\,.
\ee
Using (\ref{38}) together with the cyclicity of the trace we find 
immediately for $m^{(0)}_q=0$ 
\be\label{39}
E^\nu(x,x';G,0)=0\,.
\ee
With this and (\ref{34}) we get for $E^\nu$ of (\ref{33}) 
\be\label{39a}
E^\nu(x,x';G,m^{(0)}_q)=
m^{(0)}_qE'^\nu(x,x';G,0)+{\cal O}\left((m^{(0)}_q)^2\right)\,,
\ee
where 
\beqa\label{40A}
\lefteqn{
m^{(0)}_q E'^\nu (x,x';G,0)
}
\nn \\
&&{}= - \int d^4z\,\mathrm{Tr} \left[
S^{(q)}_F(x,z;G,0)\,m^{(0)}_q S^{(q)}_F(z,x';G,0)\gamma_5
S^{(q)}_F(x',x;G,0)\gamma^\nu 
\right.
\\
&&{}
\left. \hspace*{2.5cm}
{}
+S^{(q)}_F(x,x';G,0)\gamma_5 
S^{(q)}_F(x',z;G,0) \,m^{(0)}_q
S^{(q)}_F(z,x;G,0)\gamma^\nu\right]\,.
\nn
\eeqa
Note that only the massless propagator occurs in this expression. 
The presence of the massless propagator might potentially 
lead to infrared divergences 
when we consider the renormalisation of our amplitudes in the next 
section. But since we need to consider only the leading term in the 
light quark mass in the expansion (\ref{39a}) we can easily avoid this 
potential problem. Namely, we can replace the massless propagator 
in (\ref{40A}) by the massive propagator. The resulting expansion 
\be\label{39amitm}
E^\nu(x,x';G,m^{(0)}_q)=
m^{(0)}_qE'^\nu(x,x';G,m^{(0)}_q)+{\cal O}\left((m^{(0)}_q)^2\right)
\ee
with 
\beqa\label{40}
\lefteqn{
m^{(0)}_q E'^\nu (x,x';G,m^{(0)}_q)
}
\\
&&{}= - \int d^4z\,\mathrm{Tr} \left[
S^{(q)}_F(x,z;G,m^{(0)}_q)\,m^{(0)}_q 
S^{(q)}_F(z,x';G,m^{(0)}_q)\gamma_5
S^{(q)}_F(x',x;G,m^{(0)}_q)\gamma^\nu 
\right.
\nn \\
&&{}
\left. \hspace*{2cm}
{}
+S^{(q)}_F(x,x';G,m^{(0)}_q)\gamma_5 
S^{(q)}_F(x',z;G,m^{(0)}_q) \,m^{(0)}_q
S^{(q)}_F(z,x;G, m^{(0)}_q)\gamma^\nu\right]
\nn
\eeqa
differs from (\ref{39a}) only in terms of higher order 
in the quark mass due to (\ref{34}). 
The diagrams representing $E'^\nu$ are shown in figure \ref{figure4}. 
They correspond to a loop with a photon vertex, a pseudoscalar vertex, 
and a scalar vertex representing the quark mass insertion. 
Inserting (\ref{39amitm}) and (\ref{40}) in (\ref{32}) we see that 
$q'_\mu A^{\mu\nu}(q',q)$ is proportional to $(m^{(0)}_q)^2$, 
\beqa\label{40a}
q'_\mu A^{\mu\nu}(q',q)&=&{}
- \frac{1}{3}\int d^4x' \,d^4x\,e^{iq'x'} e^{-iqx}
\nn
\\
&&{}\hspace*{1cm}
\left[\, 2(m^{(0)}_u)^2 E'^\nu (x,x';G,m^{(0)}_u)
+(m^{(0)}_d)^2 E'^\nu (x,x';G,m^{(0)}_d)
\right]
\nn 
\\
&&{}
+{\cal O}\left((m^{(0)}_u)^3, (m^{(0)}_d)^3\right)
\,.
\eeqa

\begin{figure}[ht]
\begin{center}
\vspace*{.4cm}
\includegraphics[width=5.5cm]{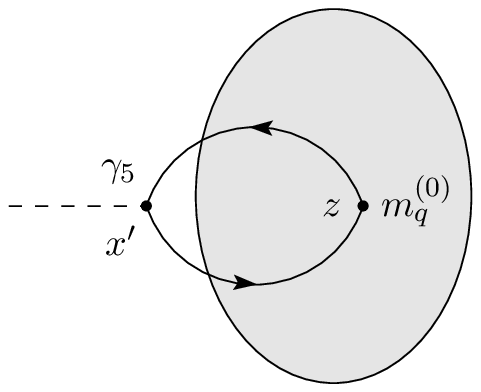}
\end{center}
\caption{Diagrammatic representation of 
$m_q^{(0)} F'(x';G,m^{(0)}_q)$ (\ref{43}) 
as a quark loop in a given gluon potential. \label{figure5}}
\end{figure}

In a similar way we can discuss the divergence of the amplitude 
$\tilde{B}^\mu(q')$ (\ref{31}). We find
\beqa\label{41}
q'_\mu\tilde{B}^\mu(q') \!\! &=&{} \!\!
iq'_\mu\frac{1}{2}\int d^4x'\,e^{iq'x'}
\mathrm{Tr} \left[\gamma^\mu\gamma_5 S^{(u)}_F(x',x';G,m^{(0)}_u)
-\gamma^\mu\gamma_5 S^{(d)}_F(x',x';G,m^{(0)}_d)\right]
\nonumber\\
&=&{} \!\!
-i\int d^4x'\,e^{iq'x'}
\left\{m^{(0)}_u \mathrm{Tr}\left[\gamma_5S^{(u)}_F(x',x';G,m^{(0)}_u)\right]
\right.
\nn \\
&&{} \hspace*{2.6cm} \left.
{}
-m^{(0)}_d \mathrm{Tr}\left[\gamma_5 S^{(d)}_F(x',x';G,m^{(0)}_d)\right]\right\}
\,.
\eeqa
Note that in the isospin symmetry limit, that is for 
$m^{(0)}_u=m^{(0)}_d$, we have $q'_\mu\tilde{B}^\mu(q')=0$. 
But in reality the light quark masses are small but quite different, 
see \cite{Gasser:1982ap} and below. 
For the trace part of the integrand in (\ref{41}) we easily find 
with (\ref{34})-(\ref{38}) for $q=u,d$ 
\beqa\label{42}
F(x';G,m^{(0)}_q)&=&{}
\mathrm{Tr}\left[\gamma_5 S^{(q)}_F(x',x';G,m^{(0)}_q)\right] 
\nn\\
&=&{}
m^{(0)}_q F'(x';G,0)+{\cal O}\left((m^{(0)}_q)^2\right)\,,
\eeqa
and 
\be
\label{43v}
m^{(0)}_q F'(x';G,0)=-\int d^4z \,
\mathrm{Tr}\left[\gamma_5 S^{(q)}_F(x',z;G,0) \,
m^{(0)}_q S^{(q)}_F(z,x';G,0)\right]\,.
\ee
Again we find it convenient to replace the massless propagator 
in this expression by the propagator for massive quarks as 
we did from (\ref{40A}) to (\ref{40}). Hence we write 
\be\label{42mitm}
F(x';G,m^{(0)}_q) =
m^{(0)}_q F'(x';G,m^{(0)}_q)+{\cal O}\left((m^{(0)}_q)^2\right)
\ee
with 
\be
\label{43}
m^{(0)}_q F'(x';G,m^{(0)}_q)=-\int d^4z \,
\mathrm{Tr}\left[\gamma_5 S^{(q)}_F(x',z;G,m^{(0)}_q)\,
m^{(0)}_q S^{(q)}_F(z,x';G,m^{(0)}_q)\right]\,,
\ee
which differs from (\ref{43v}) only by terms of higher order 
in $m_q^{(0)}$. 
The diagram corresponding to (\ref{43}) is shown in figure \ref{figure5}. 
We have a quark loop with one pseudoscalar and one scalar vertex, 
and the latter is again given by a quark mass insertion.  
Inserting (\ref{42mitm}) and (\ref{43}) in (\ref{41}) we see that also 
$q'_\mu\tilde{B}^\mu(q')$ is proportional to the square of the light 
quark masses, 
\beqa\label{43a}
q'_\mu\tilde{B}^\mu(q')&=&{} 
-i\int d^4x'\,e^{iq'x'}
\left[(m^{(0)}_u)^2 F'(x';G,m^{(0)}_u)
-(m^{(0)}_d)^2 F'(x';G,m^{(0)}_d)
\right] 
\nonumber\\
&&{}
{}+{\cal O}\left((m^{(0)}_u)^3,(m^{(0)}_d)^3\right) \,.
\eeqa

As a final point in this section we discuss the question of possible anomalous 
contributions \cite{Adler:1969gk,Bardeen:1969md,Bell:1969ts, Wess:1971yu} 
in the divergence relations (\ref{32}) 
and (\ref{41}). We are dealing here with the divergence of axial vector 
currents in an external vector (here:\ gluon) field and this is precisely the 
case studied explicitly in \cite{Bardeen:1969md}. 
The electromagnetic part of the anomaly 
is not relevant for us here since in our reactions (\ref{5}) and (\ref{16}) 
only one photon is involved. The gluon anomaly on the other hand is relevant 
for us. The divergence of the axial vector current for one quark flavour reads 
\be\label{44}
\partial_\mu\bar{q}(x)\gamma^\mu\gamma_5q(x)
=2im^{(0)}_q\bar{q}(x)\gamma_5q(x)+\frac{(g^{(0)})^2}{32 \pi^2}\,
\epsilon_{\mu\nu\rho\sigma}G^{\mu\nu}(x)G^{\rho\sigma}(x)
\,,
\ee
where we use the convention $\epsilon_{0123}=1$. 
The anomalous gluonic part of the divergence of the axial vector current in 
(\ref{44}) is, however, independent of the quark mass. Thus the anomalous 
gluonic pieces cancel in the divergence of the axial vector current $A^3_\mu$ 
of (\ref{13a}) since 
\be\label{45}
A^3_\mu(x)=
\frac{1}{2}\bar{u}(x)\gamma_\mu\gamma_5 u(x)
-\frac{1}{2}\bar{d}(x)\gamma_\mu\gamma_5 d(x) 
\,.
\ee

\section{Renormalisation}
\label{sec:renormalisation}

So far our formulae are expressed in terms of bare quantities. 
In the present section we want to consider the renormalisation 
procedure for the amplitudes obtained above. 

We use (\ref{40a}) to obtain from (\ref{26}) 
\beqa\label{46}
\lefteqn{
q'_\mu{\cal M}^{(a)\mu\nu}_{s's}(A^3;q',p,q)
=
\langle U_{s's}(p',p)q'_\mu A^{\mu\nu}(q',q)\rangle_G
}
\nonumber\\
&=&{}
-\frac{1}{3}\int d^4x' \, d^4 x\,e^{iq'x'} e^{-iqx}
\\
&&{}
\left\langle U_{s's}(p',p)\left[ \,2(m^{(0)}_u)^2 E'^\nu(x,x';G,m^{(0)}_u)
+(m^{(0)}_d)^2 E'^\nu(x,x';G,m^{(0)}_d) \right]\right\rangle_G\,.
\nn
\eeqa
Here and in the following terms of cubic or higher order in the 
light quark masses are neglected. 
Using the methods described in appendix \ref{sec:appA} 
we can show that this expression can also be obtained as the 
contribution of diagram class (a) contained in the 
following correlation function, 
\beqa\label{47}
\lefteqn{
(2\pi)^4\delta^{(4)}(p'+q'-p-q)\,
q'_\mu{\cal M}^{(a)\mu\nu}_{s's}(A^3;q',p,q)
}
\nn \\
&=&{}\!\!
- \frac{1}{2 \pi m_p} 
\int d^4x'\, d^4x \, d^4z\, e^{iq'x'} e^{-iqx}
\\
&&{} \hspace*{0.5cm} {}
\langle p(p',s')| \, \mathrm{T}^* \left[ 
\left(m^{(0)}_u\bar{u}(x')\gamma_5
u(x')\right) 
\left(m^{(0)}_u\bar{u}(z)u(z)\right) 
\right.
\nonumber\\  
&&{}\hspace*{2.5cm} {}
\left. \left. {}
-\left(m^{(0)}_d\bar{d}(x')\gamma_5d(x')\right)
\left(m^{(0)}_d\bar{d}(z)d(z)\right)
\right] 
J^\nu(x) \,
|p(p,s\rangle \,
\right|_{(a)} 
\,,
\nn 
\eeqa
see also figure \ref{figure6} below. The subscript $(a)$ in 
this expression indicates that only the diagrams of type (a) 
of the correlation function in the integrand are taken into account. 

We can obtain the divergence of ${\cal M}^{(b)\mu \nu}$ in a 
completely analogous way. We get from 
(\ref{27}),  (\ref{41}), (\ref{42mitm}) and (\ref{43a}) 
\beqa\label{54f}
\lefteqn{
q'_\mu{\cal M}^{(b)\mu\nu}_{s's}(A^3;q',p,q)
=
\langle U_{s's}(p',p)q'_\mu\tilde{B}^\mu(q')B^\nu(q)\rangle_G
}
\nonumber\\
&=&{} \!\!
-i\int d^4x'\,e^{iq'x'}
\\ 
&&{}
\left\langle U_{s's}(p',p)
\left[(m^{(0)}_u)^2 F'(x';G,m^{(0)}_u)
-(m^{(0)}_d)^2 F'(x';G,m^{(0)}_d)\right] B^\nu(q)\right\rangle_G \,.
\nn 
\eeqa
We can then relate this expression to the correlation function on the 
r.h.s.\ of (\ref{47}) involving pseudoscalar, scalar and vector currents, 
but now taking into account only the diagrams of type (b). 

It is well known (see for instance \cite{26a}) that the quark mass and 
the scalar and pseudoscalar currents have the same renormalisation 
constant $Z_{mq}$. We have for the masses
\be\label{48}
m^R_q=m^{(0)}_q Z^{-1}_{mq} \quad\quad\quad 
(q=u,d)\,,
\ee
where for definiteness we choose $m^R_q$ to be the renormalised 
quark masses in the $\overline{MS}$ scheme at renormalisation 
point $\mu=2\,\mbox{GeV}$, see \cite{Eidelman:2004wy}. 
The corresponding renormalised scalar and pseudoscalar currents are 
\beqa\label{49}
(\bar{q}(x)q(x))^R
&=&{}
Z_{mq} \, \bar{q}(x)q(x),
\nonumber\\
(\bar{q}(x)\gamma_5q(x))^R
&=&{}
Z_{mq} \, \bar{q}(x)\gamma_5q(x)
\quad\quad\quad (q=u,d)\,.
\eeqa
We have thus
\beqa\label{50}
m^R_q\left(\bar{q}(x)q(x)\right)^R
&=&{}
m^{(0)}_q\bar{q}(x)q(x)\,,
\nonumber\\
m^R_q\left(\bar{q}(x)\gamma_5q(x)\right)^R
&=&{}
m^{(0)}_q\bar{q}(x)\gamma_5q(x)\,.
\eeqa
We insert (\ref{50}) in (\ref{47}) and add the corresponding 
contribution for the divergence of the amplitude ${\cal M}^{(b)}$ 
to get 
\beqa\label{51}
\lefteqn{
(2\pi)^4\delta^{(4)}(p'+q'-p-q)\,
q'_\mu{\cal M}^{(a+b)\mu\nu}_{s's}(A^3;q',p,q)
}
\nn \\
&=&{}
-\frac{1}{2 \pi m_p}
\int d^4x'\, d^4x \, d^4z\,e^{iq'x'} e^{-iqx}
\\
&&{}
\left\{
\left. {}
(m^R_u)^2 \, \langle p(p',s')|\, \mathrm{T}^*
(\bar{u}(x')\gamma_5 u(x'))^R
(\bar{u}(z)u(z))^R
J^\nu(x) 
|p(p,s)\rangle\right|_{(a+b)}
\right.
\nonumber\\
&&{}
\left.
\left.
{}
- (m^R_d)^2 \, \langle p(p',s')|\,\mathrm{T}^*
(\bar{d}(x')\gamma_5 d(x'))^R
(\bar{d}(z)d(z))^R
J^\nu(x) 
|p(p,s)\rangle\right|_{(a+b)}\right\}\,.
\nn
\eeqa
Here we have used that in QCD the vector current $J^\nu(x)$ does 
not get renormalised. 
We now define ${\cal C}_{s's}^{(q)\nu}$ by 
\beqa\label{52}
\lefteqn{
\hspace*{-0.5cm} {}
(2\pi)^4\delta^{(4)}(p'+q'-p-q)\,
{\cal C}^{(q)\nu}_{s's}(q',p,q) 
}
\nn \\ 
&=&{} 
- \frac{1}{2 \pi m_p}
\int d^4x'\,d^4x \, d^4z \, e^{iq'x'} e^{-iqx}
\nn \\
&&{} 
\left.
\langle p(p',s')|\,\mathrm{T}^* (\bar{q}(x')\gamma_5q(x'))^R
( \bar{q}(z)q(z))^R 
J^\nu(x)
|p(p,s)\rangle
\right|_{(a+b)}\,.
\eeqa
The diagrams for the integrand in (\ref{52}) are shown in 
figure \ref{figure6}. 
Note that even for massless quarks we have 
${\cal C}_{s's}^{(u)\nu} \neq {\cal C}_{s's}^{(d)\nu}$, because 
$u$ and $d$ quarks contribute differently in the current $J^\nu$. 
Since ${\cal C}^{(q)\nu}$ in (\ref{52}) involves only renormalised 
quantities it should be finite. With (\ref{52}) we obtain from (\ref{51}) 
\be \label{53}
q'_\mu{\cal M}^{(a+b)\mu\nu}_{s's}(A^3;q',p,q)
= (m^R_u)^2 \, {\cal C}^{(u)\nu}_{s's}(q',p,q)
- (m^R_d)^2 \,{\cal C}^{(d)\nu}_{s's}(q',p,q)\,.
\ee
In (\ref{51}) and (\ref{53}) we have again neglected terms of 
cubic or higher order in the light quark masses. 
\begin{figure}[ht]
\begin{center}
\vspace*{.4cm}
\includegraphics[width=12.5cm]{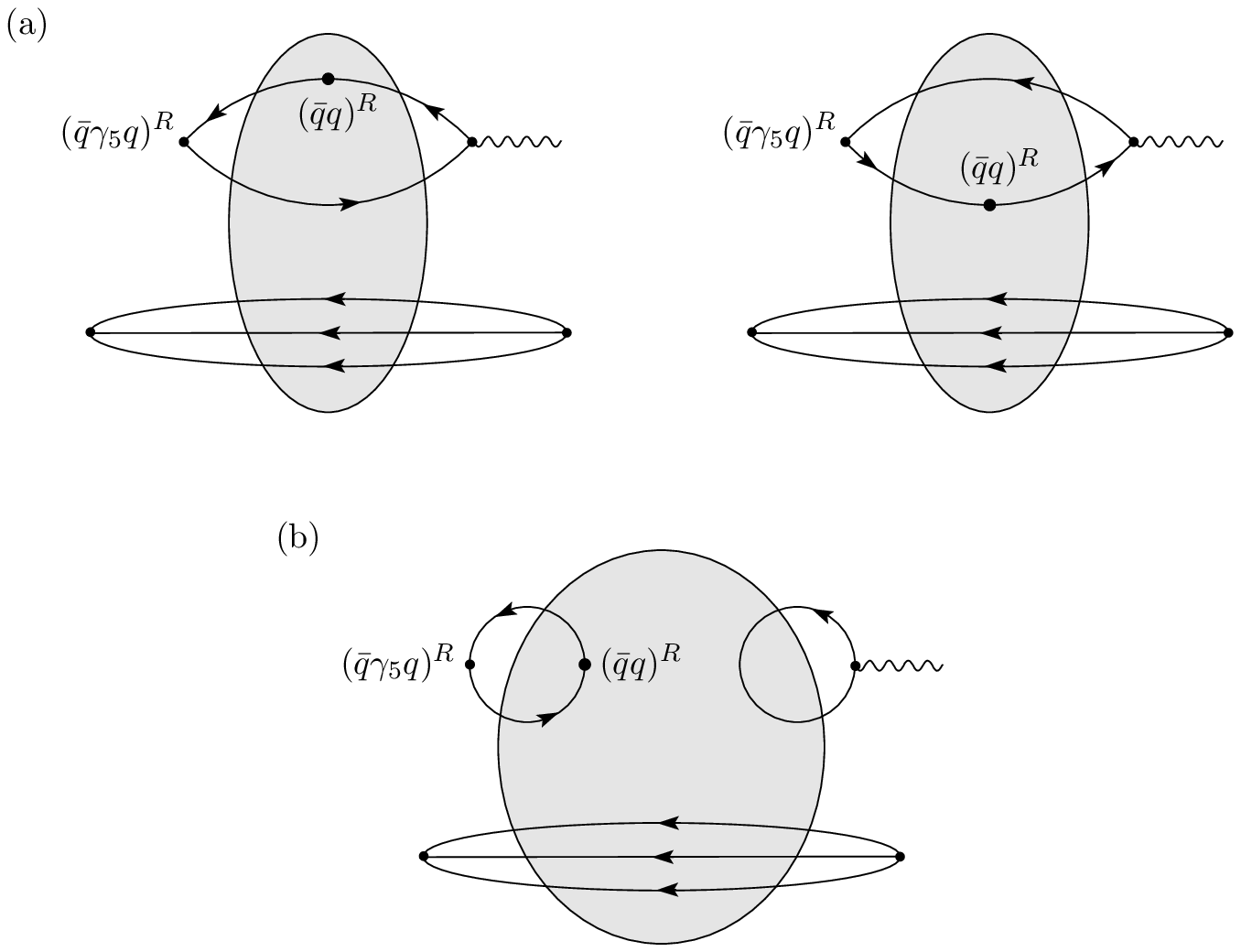}
\end{center}
\caption{The diagrams of type (a) and (b) represent the 
correlation function ${\cal C}^{(q)\nu}_{s's}$ of (\ref{52}). 
\label{figure6}}
\end{figure}

With (\ref{53}) we have shown that the divergence 
amplitude $q'_\mu{\cal M}^{(a+b)\mu\nu}_{s's}(A^3;q',p,q)$ 
is indeed proportional to the square of the renormalised light quark 
masses. 

\section{Results and conclusions}
\label{sec:results}

It is well known that the light quark masses are directly related 
to $m^2_\pi$. Indeed one finds in the chiral limit (see (8.1) of 
\cite{Gasser:1982ap}) for the average of the quark masses
\be\label{56}
\hat{m}\equiv \frac{1}{2}(m^R_u+m^R_d)
=\frac{1}{2B} \,m^2_\pi\,.
\ee
Here 
\be\label{57}
B=-\frac{2}{f^2_\pi}\langle 0|\left(\bar{u}(x)u(x)\right)^R|0\rangle 
\ee
is a hadronic constant which stays finite in the chiral limit 
$m^R_{u,d}\rightarrow 0$. 

Experimentally the light quark masses still are not too well 
known, see \cite{Eidelman:2004wy}. In the following we 
take our estimates of `central' values extracted from 
\cite{Eidelman:2004wy} and assume these masses to be 
$m_u^R \cong 3\,\mbox{MeV}$ and $m_d^R \cong 7\,\mbox{MeV}$ 
at a renormalisation scale of $2\,\mbox{GeV}$. 
This gives $\hat{m} \cong 5\,\mbox{MeV}$, and with 
$m_\pi = 135\,\mbox{MeV}$ we get $B\cong  1820 \, \mbox{MeV}$. 
The ratios of the light quark masses and the mean are then 
\be\label{59}
r_u=\frac{m^R_u}{\hat{m}}\cong  0.6 \,,\quad \quad \quad
r_d=\frac{m^R_d}{\hat{m}}\cong  1.4 \,.
\ee

Now we go back to (\ref{20}) and discuss the quark mass, 
respectively $m^2_\pi$, dependence of ${\cal M}^\nu_{s's}(\pi^0;q',p,q)$ 
in the chiral limit $m^R_{u,d}\rightarrow 0$. 
For high energies we consider only the odderon-exchange diagrams 
(a) and (b) of figure \ref{figure3} for the reasons given in section 
\ref{sec:classification}. 
We have then for the amplitudes corresponding to the sum of 
those diagrams  
\be\label{60}
{\cal M}^{(a+b)\nu}_{s's}(\pi^0;q',p,q)\equiv 
\frac{2\pi m_p\sqrt{2}}{f_\pi m^2_\pi}\,
(-q'^2+m^2_\pi) \,iq'_\mu{\cal M}^{(a+b)\mu\nu}_{s's}(A^3;q',p,q)
\ee
and obtain from (\ref{53}) with (\ref{56}), (\ref{59}) 
\beqa\label{61}
{\cal M}^{(a+b)\nu}_{s's}(\pi^0;q',p,q)
&=&{}
m^2_\pi \,\frac{\pi m_p}{f_\pi B^2\sqrt{2}}
\left[- 
r^2_u  \,i(q'^2-m^2_\pi) \,{\cal C}^{(u)\nu}_{s's}(q',p,q)
\right.
\nn \\
&&{}\hspace*{2.2cm}
\left. {}
+ r^2_d\,i(q'^2-m^2_\pi)\,{\cal C}^{(d)\nu}_{s's}(q',p,q)\right] \,.
\eeqa
The correlation functions ${\cal C}^{(u)\nu}$ and ${\cal C}^{(d)\nu}$ 
occurring in (\ref{61}) are properly renormalised. 
It is clear from their definition 
in (\ref{52}) (see also figure \ref{figure6}) that they will 
have pion poles which are just cancelled by the explicit factors 
$(q'^2-m^2_\pi)$ in (\ref{61}). Otherwise these 
functions should be finite in the chiral limit. Also $m_p$, $f_\pi$ 
and $B$ are known to approach finite values in the chiral limit, 
see for example \cite{Gasser:1982ap}. Thus, due to the explicit 
factor $m^2_\pi$ in (\ref{61}) the odderon-exchange 
amplitude for the reaction $\gamma^{(*)}p\to \pi^0 p$ 
vanishes in the chiral limit $m^2_\pi\rightarrow 0$. 
This is the main result of the present paper.

In the case of approximate chiral symmetry, as realised in Nature, 
we do not expect the odderon-exchange 
amplitude for the reaction $\gamma^{(*)}p\to \pi^0 p$ 
to vanish exactly. But from the above result we 
should expect that the approximate chiral symmetry leads to 
a strong dynamical suppression of this amplitude. It is difficult to 
assess the numerical effect of this suppression, a rough estimate has 
been given in \cite{Donnachie:2005bu}. It indicates that the effect  
of approximate chiral symmetry can modify the prediction 
(\ref{2}) of \cite{Berger:1999ca} such as to reconcile it with 
the experimental upper bound (\ref{3}) on the diffractive 
photoproduction of neutral pions. 

Our result (\ref{61}) holds for all photon virtualities $Q^2$ and momentum 
transfers $\sqrt{-t}$. In particular, it should also extend into the 
perturbative region of large $Q^2$ or large $\sqrt{-t}$. In this 
context it is worth pointing out that the result matches nicely 
the perturbative result of \cite{Braunewell:2005ct} where 
the diffractive reaction $\gamma^{(*)} p \to \eta_c \,p$ was 
considered at high energies. In that reaction perturbation theory can 
be applied because of the large scale given by the charm quark mass. 
In leading order in perturbation theory only diagrams of type (a) 
contribute to the $\gamma^{(*)} \to \eta_c$ impact factor. In 
\cite{Braunewell:2005ct} this impact factor was computed 
for an arbitrary number of gluons exchanged in the $t$-channel. 
It was found that for any number of exchanged gluons the impact 
factor, and hence the amplitude for that reaction, is linear in the quark 
mass. That agrees with the result that we find here based on general 
nonperturbative calculations, and the mechanism leading to that result 
is in fact the perturbative realisation of the one that we have described 
here in section \ref{sec:divrel}. 

\begin{figure}[ht]
\begin{center}
\vspace*{.4cm}
\includegraphics[width=14.45cm]{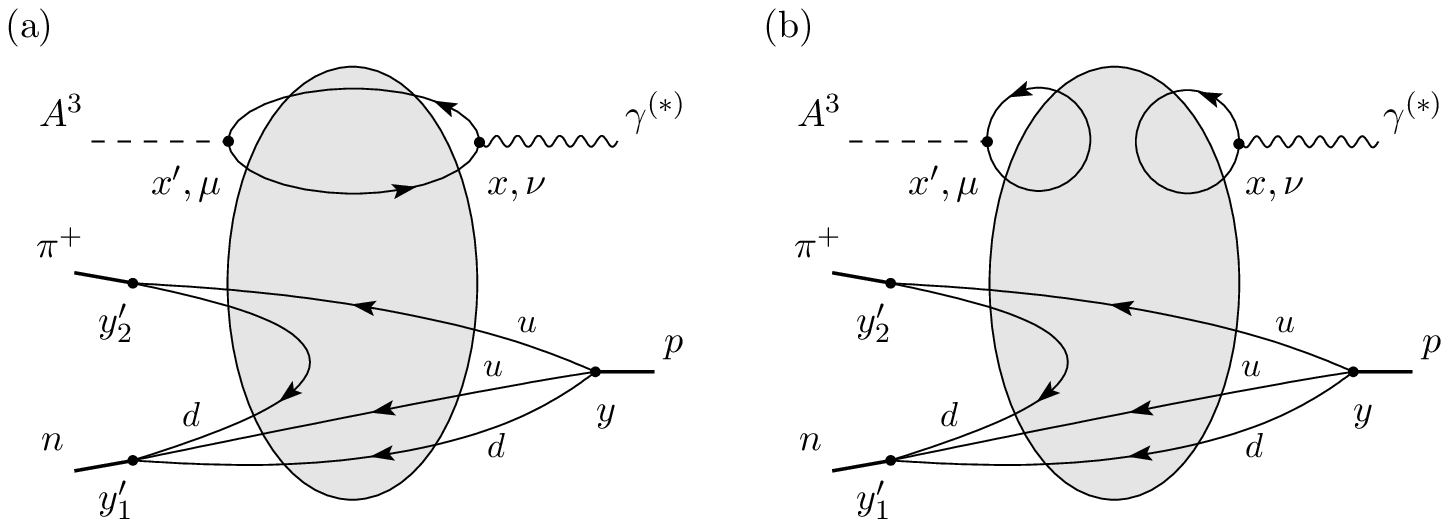}
\end{center}
\caption{Odderon exchange diagrams for 
$\gamma^* p\rightarrow A^3 n\pi^+$, reaction (\ref{64}). \label{figure7}}
\end{figure}

Let us point out that we can easily extend our result to the 
general reaction (\ref{1}) with nucleon dissociation. 
As an example we discuss in appendix \ref{sec:appB} 
the reactions 
\be\label{63}
\gamma^{(*)}(q,\nu)+p(p,s)\rightarrow\pi^0(q')+n(p'_1,s')+\pi^+(p'_2)
\ee
and
\be\label{64}
\gamma^{(*)}(q,\nu)+p(p,s)\rightarrow A^3(q',\mu)+n(p'_1,s')+\pi^+(p'_2) \,.
\ee
With the same techniques as above we find that taking the 
divergence of the odderon-exchange diagrams for (\ref{64}) 
(see figure \ref{figure7}) gives an explicit factor $m^2_\pi$ 
in the amplitude for (\ref{63}). 
That is, the odderon-exchange contribution to the reaction (\ref{63}) 
vanishes in the chiral limit. 

Finally, our findings can also be generalised to reactions of two 
real or virtual photons which will be relevant at the LHC and at a 
future ILC. Namely, it is straightforward to apply the same techniques 
to the diffractive reaction $\gamma^{(*)} + \gamma^{(*)} \to \pi^0 + X$ 
at high energy. Again we find that the amplitude for the 
odderon-exchange contribution to this process is proportional to $m_\pi^2$ 
and vanishes in the chiral limit. In the diffractive reaction 
$\gamma^{(*)} + \gamma^{(*)} \to \pi^0 + \pi^0$ the odderon-exchange 
contribution is even suppressed by a factor $m_\pi^4$ in the amplitude. 
We therefore expect the cross sections for these processes to be very 
small at high energies, independently of the odderon intercept. 

To summarise, we have studied the diffractive photo- and 
electroproduction of a neutral pion on a nucleon, 
$\gamma^{(*)} + N \to \pi^0 + X$ (reaction (\ref{1})). 
We have shown that the diagrams with multi-gluon exchange 
in the $t$-channel, that is the odderon exchange diagrams, vanish 
in the chiral limit. In the real world with approximate chiral 
symmetry these diagrams are dynamically suppressed by 
a factor $m_\pi^2$, and hence the cross section by a factor 
$m_\pi^4$. At high energies the other types of diagrams for 
reaction (\ref{1}) (see figure 2 of \cite{Ewerz:2004vf}) 
are expected to be suppressed by inverse powers of the c.\,m.\ energy. 
Thus we have as a firm prediction of QCD that the cross sections 
for the reactions (\ref{1}) should be very small at high energies 
compared to cross sections for reactions in which pomeron 
exchange is allowed like for instance $\gamma^{(*)} + N \to \rho^0 +X$. 

\section*{Acknowledgements}

The authors are grateful to S.\ Braunewell, A.\ Donnachie and 
H.\,G.\ Dosch for helpful discussions. 

\begin{appendix}
\numberwithin{equation}{section}

\section{Functional methods}
\label{sec:appA}

Here we give a short outline of the functional method introduced 
in \cite{Ewerz:2004vf} which is used to derive (\ref{26}) 
and (\ref{27}). We start from (\ref{18}) and use the LSZ reduction 
formula \cite{Lehmann:1954rq} to represent the amplitude as an integral 
over a Green's function. The latter is then represented as a functional 
integral and the quark degrees of freedom are integrated out, 
giving rise to nonperturbative quark propagators in a given gluon 
potential. This leads to a number of nonperturbative diagrams distinguished 
by the topology of the quark-line skeleton. 

In detail, let $\psi_p(x)$ be a suitable interpolating field operator for 
the proton. We can take $\psi_p$ to contain three quark fields, 
\be\label{A.1}
\psi_p(x)=\Gamma_{\alpha\beta\gamma}
u_{\alpha}(x)u_{\beta}(x)d_{\gamma}(x) \,,
\ee
and accordingly
\be\label{A.2}
\bar{\psi}_p(x)=\bar{\Gamma}_{\alpha\beta\gamma}
\bar{d}_{\gamma}(x)\bar{u}_{\beta}(x)\bar{u}_{\alpha}(x)\,,
\ee
where $\Gamma_{\alpha\beta\gamma}$ and 
$\bar{\Gamma}_{\alpha\beta\gamma}$ are coefficient matrices and 
$\alpha,\beta,\gamma$ summarise Dirac and colour indices. 
Explicit constructions of field operators $\psi_p(x)$ of the type (\ref{A.1}) 
can be found in \cite{Chung:wm,Chung:cc,Ioffe:kw}. 
Let $Z_p$ be the proton's wave function renormalisation constant defined by 
\be\label{A.3}
\left\langle0|\psi_p(x)|p(p,s)\right\rangle
=\sqrt{Z_p} \, e^{-ipx}u_s(p)\,.
\ee
As explained above, the amplitude (\ref{17}) can be represented in terms 
of a functional integral
\beqa\label{A.4}
\lefteqn{
{\cal M}^{\mu\nu}_{s's}(A^3;q',p,q)=
-\frac{i}{2\pi m_p Z_p}
\int d^4 x' \, d^4x \, d^4y\,e^{iq'x'} e^{-iqx} e^{-ipy}
}
\nonumber\\
&&{}
\left\{\bar{u}_{s'}(p')
(-i\rightslash_{y'}+m_p)
\frac{1}{{\cal Z}}\int{\cal D}(G,q,\bar{q})\,
\bar{\psi}(x')\gamma^\mu\gamma_5\mathbf{T}_3\psi(x')
\bar{\psi}(x)\gamma^\nu\mathbf{Q}\psi(x)
\right.
\nonumber\\
&&{}
\left. \left.
\psi_p(y')\bar{\psi}_p(y)
\exp\left[i\int d^4z \,{\cal L}_{\rm QCD}(z)\right]
(i\leftslash_y+m_p)u_s(p)
\right\} \right|_{y'=0}\,.
\eeqa
Gauge fixing and Fadeev-Popov terms are implied and not written out 
explicitly.  

Now we can integrate out the quark degrees of freedom in the functional 
integral since ${\cal L}_{\rm QCD}$ is bilinear in the quark fields. 
This leads to a purely gluonic functional integral where all explicit 
quark fields in (\ref{A.4}) are contracted out. As in (A.6) 
of \cite{Ewerz:2004vf} the contraction of 
two quark fields of flavour $q$ is defined as 
\be\label{A.5}
\wick{1}{<1 q(x)>1{\overline{q}}(y)=\frac{1}{i}S^{(q)}_F(x,y;G)} \,,
\ee
that is, as the quark propagator in the given gluon potential $G$. 
The propagator (\ref{A.5}) satisfies (see (16) and appendix A 
of \cite{Ewerz:2004vf})
\beqa\label{A.6}
(i\gamma^{\mu}D_{\mu}-m^{(0)}_q)S^{(q)}_F(x,y;G)
&=&{}
\left( i \! \slash{\partial}_x - g^{(0)} \! \slash{G}^a \frac{\lambda_a}{2} 
- m_q^{(0)} \right) S^{(q)}_F(x,y;G)
\nonumber\\
&=&{} 
-\delta^{(4)}(x-y)
\eeqa
and
\beqa\label{A.7}
S_F^{(q)} (x,y;G) ( i \leftDslash_y + m_q^{(0)} ) 
&=&{}
 S_F^{(q)} (x,y;G) \left( i \leftslash_y + \, g^{(0)}\! \slash{G}^a 
\frac{\lambda_a}{2} + m_q^{(0)} \right) 
\nonumber\\
&=&{}
\delta^{(4)} (x-y) \,,
\eeqa
where $g^{(0)}$ is the bare coupling parameter and $m^{(0)}_q$ are 
the bare quark masses. 
Performing these contractions for the functional integral in (\ref{A.4}) 
we get a number of terms, (a) to (g), in analogy to figure 2 
of \cite{Ewerz:2004vf}. The diagrams (a) and (b) are shown in figure 
\ref{figure3}. The shaded blobs correspond to the functional integral 
$\langle\cdot\rangle_G$. For any functional $F[G]$ we define
\beqa\label{A.8}
\left\langle F[G]\right\rangle_G&=&\frac{1}{\mathcal{Z}^{\prime}}
\int \mathcal{D}(G)F[G] \,
\prod\limits_q \det\left[-i( i \gamma^{\lambda}
D_{\lambda}-m^{(0)}_q+i\epsilon)\right]\nonumber\\
&&{}
\exp\left[-i\int d^4x \, \frac{1}{2}\mathrm{Tr} 
\left( G_{\lambda\rho}(x)G^{\lambda\rho}(x)\right)
\right]\,,
\eeqa
where $G_{\lambda \rho}$ is the matrix-valued gluon field strength tensor 
and ${\cal Z}'$ is the normalisation factor obtained from the condition 
\be\label{Q.9}
\langle 1\rangle_G=1\,.
\ee
In (\ref{A.8}) the fermion determinant is included, gauge fixing and 
Faddeev-Popov terms are implied. All quantities in (\ref{A.8}) are the 
unrenormalised ones. The product over $q$ runs over all quark flavours. 

In the diagram figure \ref{figure3}a the quarks of the axial vector and 
the electromagnetic current in (\ref{A.4}) are contracted with 
each other as are the fields of the proton operators. 
This leads to ${\cal M}^{(a)\mu\nu}$ (\ref{26}). 
The contraction of the quark fields in $\psi_p$ with those in 
$\bar{\psi}_p$ gives 
\beqa\label{A.10}
\wick{1}{<1\psi_p(y^{\prime})>1{\overline{\psi}}_p(y)}
&=&
\Gamma_{\alpha^{\prime}\beta^{\prime}\gamma^{\prime}}
\bar{\Gamma}_{\alpha\beta\gamma} \,
\frac{1}{i}S^{(d)}_{F\gamma^{\prime}\gamma}(y^{\prime},y;G)
\nn \\
&&{}
\biggl\{\frac{1}{i}S^{(u)}_{F\beta^{\prime}\beta}(y^{\prime},y;G)
\frac{1}{i}S^{(u)}_{F\alpha^{\prime}\alpha}(y^{\prime},y;G)-
(\alpha\leftrightarrow \beta)\biggr\}
\,.
\eeqa
The quantity $U_{s's}(p',p)$ in (\ref{26}) and (\ref{27}) representing the 
lower part of the diagrams of figures \ref{figure3}a and \ref{figure3}b, 
that is the scattering of the proton in the fixed gluon potential, 
is defined as 
\beqa\label{A.11}
U_{s's}(p',p)&=&
{}-\frac{i}{2\pi m_pZ_p}\int d^4y\,e^{-ipy}
\left[\bar{u}_{s'}(p')
(-i\rightslash_{y'}+m_p)
\right.
\nonumber\\
&&{}
\left. \left.
\wick{1}{<1 \psi_p(y')>1{\overline{\psi}}_p(y)}
(i\leftslash_y+m_p)u_s(p)
\right]\right|_{y'=0}\,.
\eeqa
In the matrix element ${\cal M}^{(b)\mu\nu}$ the quarks of the 
axial vector current in (\ref{A.4}) are contracted among themselves as are the 
quarks of the electromagnetic current. The contraction of the proton fields 
is as for ${\cal M}^{(a)\mu\nu}$. 

The matrix elements ${\cal M}^{(c)\mu\nu}_{s's}$ to 
${\cal M}^{(g)\mu\nu}_{s's}$ in (\ref{25}) are the analogues of the 
diagram classes of figure 2c to 2g in \cite{Ewerz:2004vf} 
with the electromagnetic 
current for the photon $\gamma(\mu)$ replaced by the axial vector current. 
As discussed in \cite{Ewerz:2004vf} these diagrams do not correspond 
to multi-gluon exchange in the $t$-channel. They are  expected to give 
only small contributions at high energies. Typically we expect for them 
Regge behaviour corresponding to the exchange of meson trajectories or 
fermion trajectories. 

\section{Reactions with nucleon dissociation}
\label{sec:appB}

Here we discuss the reactions (\ref{63}) and (\ref{64}). With the charge 
symmetry relation we get the interpolating field for the neutron 
from (\ref{A.1}) as 
\be\label{B.1}
\psi_n(x)=-\Gamma_{\alpha\beta\gamma}\,
d_\alpha(x)d_\beta(x)u_\gamma(x) \,.
\ee
According to (\ref{15}) and (\ref{9}) we have for the interpolating 
$\pi^+$-field 
\be\label{B.2}
\phi_+(x)=\frac{1}{f_\pi m^2_\pi}\,\partial_\lambda
\left[A^{1\lambda}(x)-iA^{2\lambda}(x)\right]
=\frac{1}{f_\pi m^2_\pi}\,\partial_\lambda
\left[\bar{d}(x)\gamma^\lambda\gamma_5 u(x)\right]\,.
\ee
We define the amplitudes for (\ref{63}) and (\ref{64}) as follows: 
\beqa\label{B.3}
\lefteqn{
(2\pi)^4\delta^{(4)}(p'_1+p'_2+q'-p-q)\,
{\cal M}^\nu_{s's}(\pi^0,n,\pi^+;p'_2,q',p,q)
}
\\
&=&{} -i \int d^4x'\,d^4x\,e^{iq'x'} e^{-iqx}
\left(\Box_{x'}+m^2_\pi\right) \langle\pi^+(p'_2),n(p'_1,s')|\,
\mathrm{T}^*\phi^3(x')J^\nu(x)|p(p,s)\rangle \,,
\nonumber\\
\label{B.4}
\lefteqn{
(2\pi)^4\delta^{(4)}(p'_1+p'_2+q'-p-q) \,
{\cal M}^{\mu\nu}_{s's}(A^3,n,\pi^+;p'_2,q',p,q)
}
\nonumber\\
&=&{}
\frac{i}{2\pi m_p}\int d^4x' \,d^4x\,e^{iq'x'} e^{-iqx}
\langle\pi^+(p'_2),n(p'_1,s')|\,\mathrm{T}^*A^{3\mu}(x')J^\nu(x)
|p(p,s)\rangle \,.
\eeqa
As in (\ref{20}) we obtain here from PCAC (\ref{15}) 
\be\label{B.5}
{\cal M}^\nu_{s's}(\pi^0,n,\pi^+;p'_2,q',p,q)=
\frac{2\pi m_p\sqrt{2}}{f_\pi m^2_\pi}\,
(-q'^2+m^2_\pi)\,
iq'_\mu{\cal M}^{\mu\nu}_{s's}(A^3,n,\pi^+;p'_2, q',p,q)\,.
\ee

With the LSZ reduction formula \cite{Lehmann:1954rq} we get 
from (\ref{B.4}) 
\beqa\label{B.6}
\lefteqn{
{\cal M}^{\mu\nu}_{s's}(A^3,n,\pi^+; p'_2,q',p,q)
}
\nn \\
&=&{}
\frac{1}{2\pi m_p Z_p}\int d^4x' \,d^4x\,d^4y'_2 \,d^4y \, 
e^{iq'x'} e^{-iqx} e^{ip'_2y'_2} e^{-ipy}
\nonumber\\
&&{}
\left[ \left(\Box_{y'_2}+m^2_\pi\right)\bar{u}_{s'}(p'_1)
(-i\rightslash_{y'_1} \!+m_n)
\langle 0 |\mathrm{T}^* \phi_+(y'_2)\psi_n(y'_1)A^{3\mu}(x')J^\nu(x)
\bar{\psi}_p(y) |0\rangle 
\right.
\nonumber\\
&&{}
\left. \left. (i\leftslash_y \!+m_p)u_s(p)\right] \right|_{y'_1=0}
\,.
\eeqa
The next step is to represent the Green's function in (\ref{B.6}) as a 
functional integral and to integrate out the quark degrees of freedom. 
This leads to a number of nonperturbative diagrams. The important 
ones for us, that is the odderon exchange diagrams, are shown in 
figure \ref{figure7}. Again there are diagrams of type (a) and (b). 
The corresponding analytic expressions are as follows 
\beqa\label{B.7}
{\cal M}^{(a)\mu\nu}_{s's}(A^3,n,\pi^+;p'_2,q',p,q)
&=&{}
\left\langle\tilde{U}_{s's}(p'_1,p'_2,p)A^{\mu\nu}(q',q)
\right\rangle_G\,,\\
{\cal M}^{(b)\mu\nu}_{s's}(A^3,n,\pi^+;p'_2,q',p,q)
&=&{}\label{B.8}
\left\langle\tilde{U}_{s's}(p'_1,p'_2,p)\tilde{B}^\mu(q')B^\nu(q)
\right\rangle_G\,.
\eeqa
Here we define 
\beqa\label{B.9}
\lefteqn{
\tilde{U}_{s's}(p'_1,p'_2,p)
=\frac{1}{2\pi m_p Z_p} \int d^4 y'_2 \,d^4y \,e^{ip'_2y'_2} e^{-ipy} 
}
\\
&&{}
\left.
\left[\left(\Box_{y'_2}+m^2_\pi\right)\bar{u}_{s'}(p'_1)
(-i\rightslash_{y'_1}+m_n)
\wick{1}{<1\phi_+(y'_2)\psi_n(y'_1)>1{\overline{\psi}}_p}(y)
(i\leftslash_y+m_p)u_s(p)\right] 
\right|_{y'_1=0} \,,
\nonumber
\eeqa
where 
\beqa
\label{B.10}
\lefteqn{
\wick{1}{<1\phi_+(y'_2)\psi_n(y'_1)>1{\overline{\psi}}_p}(y)=
\frac{1}{f_\pi m^2_\pi}\,\frac{\partial}{\partial y'^\lambda_2}\,
\Gamma_{\alpha'\beta'\gamma'}\bar{\Gamma}_{\alpha\beta\gamma}
(\gamma^\lambda\gamma_5)_{\delta\delta'}
}
\\
&&{}
\hspace*{1cm}
\left[\frac{1}{i}S^{(d)}_{F\beta'\delta}(y'_1,y'_2;G)
\frac{1}{i}S^{(d)}_{F\alpha'\gamma}(y'_1,y;G)
-\frac{1}{i}S^{(d)}_{F\alpha'\delta}(y'_1,y'_2;G)
\frac{1}{i}S^{(d)}_{F\beta'\gamma}(y'_1,y;G)
\right]
\nonumber\\
&&{}
\hspace*{1cm}
\left[\frac{1}{i}S^{(u)}_{F\delta'\beta}(y'_2,y;G)
\frac{1}{i}S^{(u)}_{F\gamma'\alpha}(y'_1,y;G)
-\frac{1}{i}S^{(u)}_{F\delta'\alpha}(y'_2,y;G)
\frac{1}{i}S^{(u)}_{F\gamma'\beta}(y'_1,y;G)\right]
\,.
\nonumber
\eeqa
The expressions (\ref{B.7}) and (\ref{B.8}) are completely analogous 
to (\ref{26}) and (\ref{27}), respectively. They contain the same 
quantities $A^{\mu\nu}(q',q)$ (\ref{28}), $B^\nu(q)$ (\ref{30}) and 
$\tilde{B}^\mu(q')$ (\ref{31}). The $q'$-dependence is contained only in 
$A^{\mu\nu}(q',q)$ and $\tilde{B}^\mu(q')$. The discussion of the divergences 
$q'_\mu{\cal M}^{(a,b)\mu\nu}(A^3,n,\pi^+;p'_2,q',p,q)$ is, therefore, 
the same as in sections \ref{sec:classification} to \ref{sec:renormalisation} 
for reaction (\ref{16}). The final result is 
that also the odderon-exchange amplitudes for the proton break-up 
reaction (\ref{63}) are proportional to $m^2_\pi$,  
\be\label{B.11}
{\cal M}^{(a,b)\nu}_{s's}(\pi^0,n,\pi^+;p'_2,q',p,q)\propto m^2_\pi \,,
\ee
and, therefore, vanish in the chiral limit. 

The generalisation of this result to the odderon-exchange amplitudes of 
other nucleon break-up reactions is straightforward.

\end{appendix}

\end{document}